\documentclass[twocolumn,english,aps,pra,superscriptaddress,showpacs,eqsecnum]{revtex4}
\usepackage[T1]{fontenc}
\usepackage[latin1]{inputenc}
\usepackage{amsmath}
\usepackage{graphicx}

\makeatletter
\usepackage{graphicx,psfrag,amsmath,amssymb,float}
\usepackage{epstopdf}
\usepackage{color,times}
\input{epsf}
\usepackage{pslatex}
\usepackage{amsmath}
\usepackage{epstopdf}
\usepackage{graphicx}
\usepackage{amssymb,epsfig}

\usepackage{babel}
\makeatother
\begin{document}

\title{Quantum channels in random spin chains}

\author{Jos\'e A. Hoyos}

\email{hoyosj@umr.edu}

\affiliation{Department of Physics, University of Missouri-Rolla, Rolla, Missouri
65409, USA.}

\author{Gustavo Rigolin}

\email{rigolin@ifi.unicamp.br}

\affiliation{Instituto de F\'{\i}sica Gleb Wataghin, Universidade Estadual de
Campinas, Caixa Postal 6165, CEP 13083-970, Campinas, S\~ao Paulo,
Brazil.}

\begin{abstract}
We study the entanglement between pairs of qubits in a random antiferromagnetic
spin-1/2 chain at zero temperature. We show that some very distant
pairs of qubits are highly entangled, being almost pure Bell states.
Furthermore, the probability to obtain such spin pairs is proportional
to the chain disorder strength and inversely proportional to the square
of their separation.
\end{abstract}


\maketitle

\section{Introduction}

The concept of entanglement between two systems dates back to the
birth of quantum mechanics~\cite{epr,schroedinger}. In a certain
sense, entanglement can be associated with the counterintuitive nonlocal
behavior of two quantum systems: namely, the appearance of nonlocal
correlations between them even after their mutual interaction is switched
off. Such a feature has puzzled our minds, and explorations of that
concept have opened new fields of research such as quantum computation
and quantum information~\cite{livro-do-nielsen}. In particular,
many useful quantum communication protocols such as superdense coding~\cite{wiesner},
teleportation~\cite{bennett}, and quantum state sharing~\cite{hillery,li}
require entangled states (quantum channels) as a key ingredient for
their successful implementation. 

In view of that, the search for physical systems where entangled states
can be found turned to be of utmost importance not only for academic
reasons, but also for practical implementation of the new technology
arising from these new fields of research. Promising physical systems
in which entanglement may be abundant are quantum spin chains. For
a long time one has intensively studied such systems mainly because
of their interesting magnetic properties. Therefore, a great amount
of knowledge about the physics of such many-body systems was obtained,
paving the way to an investigation of the entanglement properties
of spin chains~\cite{osborne}. A fruitful result of this investigation,
for example, is the discovery of the role of entanglement in quantum
phase transitions. At the critical point, it is known that the bipartite
entanglement entropy~\cite{footnote1} of the ground state diverges
logarithmically with the size of the subsystem. In addition to this,
it was shown that the bipartite entanglement is proportional to a
quantity called central charge, which dictates the conformal field
theory universality class of the chain~\cite{holzheya-larsena-wilczek}. 

Almost all previous studies dealt with ordered spin chains. Recently,
nevertheless, it was shown that the bipartite entanglement of a random
antiferromagnetic (AF) Heisenberg spin-1/2 chain also grows logarithmically
with the size of the subsystem~\cite{refael-moore,laflorencie}.
A similar result was also derived for a random transverse-field Ising
chain at the critical point. In both cases the `effective' central
charge is less than the one calculated for their clean and ordered
counterparts~\cite{refael-moore}. In other words, bipartite entanglement
entropy is decreased when disorder is added into the system. In view
of such a result, one may naively think that disorder is an undesired
feature in spin chains when one is concerned with implementation of
any quantum information protocol. Here we show, interestingly, that
disorder provides a ground state with remarkable entanglement properties
which can be properly harnessed to create almost maximally entangled
Bell states, the ultimate quantum channel needed in many quantum communication
protocols. 

In this work we study the entanglement properties of long-distance
pairs of qubits in the random AF spin-1/2 chain. For that purpose
we determine for some suitable pairs in the chain how close they are
to a Bell state; i.e., we compute the fidelity of those pairs with
a singlet. In addition, we exhaustively characterize their entanglement
properties computing their negativity~\cite{werner}, logarithmic
negativity~\cite{werner}, and concurrence~\cite{concurrence}.
Moreover, we relate all these quantities to each other. We show that
some rare pairs display strikingly high fidelities, qualifying them
as good quantum channels. Finally, we show that the fidelity distribution
of such rare pairs does not depend on the size of the chain and is
proportional only to the disorder strength. 

The remainder of this paper is organized as follows: We introduce
the model, compute the fidelity, and relate it to the negativity,
logarithmic negativity, and concurrence in Sec.~\ref{sec:The-model}.
In Sec.~\ref{sec:Numerical-results}, we present our numerical data,
quantifying the interesting pairwise entanglement properties of the
disordered system, and compare it with the ordered and clean ones.
Finally, we conclude and discuss the relevance of our results in Sec.~\ref{sec:Conclusions}.

\section{The model and its pairwise entanglement\label{sec:The-model}}

In this section we introduce the model and compute its pairwise entanglement
as a function of the spin-spin correlation functions.

\subsection{The model}

Our starting point is the random XXZ AF spin-1/2 chain whose Hamiltonian
reads \begin{equation}
H=\sum_{i}J_{i}\left(S_{i}^{x}S_{i+1}^{x}+S_{i}^{y}S_{i+1}^{y}+\Delta_{i}S_{i}^{z}S_{i+1}^{z}\right),\label{eq:hamiltonian}\end{equation}
 where $S_{i}$ is the usual spin-1/2 operator related to site $i$
and $J_{i}>0$ and $\Delta_{i}$ are independent random variables.
In all the following calculations we use $\Delta_{i}=0$, $\forall\, i$.
However, we first argue that our results are valid for all models
satisfying the condition $-1/2<\Delta_{i}\leq1$. It was shown that
weak anisotropy ($-1/2<\Delta_{i}<1$) renormalizes to zero in the
low-energy limit, being thus irrelevant~\cite{doty-fisher,fisher94-xxz}.
Furthermore, the isotropic point $\Delta_{i}=1$, $\forall\, i$,
is governed by the same infinite-randomness fixed point (IRFP) that
governs the point $\Delta_{i}=0$; i.e., those systems quantitatively
share the same long-length ground state properties~\cite{fisher94-xxz}.
Thus, from now on it is sufficient to consider only the case where
$\Delta_{i}=0$. This choice allows us to map the Hamiltonian (\ref{eq:hamiltonian})
to a free fermion one through a Jordan-Wigner transformation. Then,
we can exactly diagonalize large chains using standard subroutines~\cite{lieb-schultz-mattis}. 

The phase diagram of the clean counterpart of this system ($J_{i}=J$
and $\Delta_{i}=\Delta$) is well known~\cite{baxter}. For $-1<\Delta\leq1$
the system is a critical Tomonaga-Luttinger spin liquid~\cite{haldane}.
For $\Delta>1$, a spin gap opens and the system belongs to the Ising
antiferromagnet universality class, and for $\Delta\leq-1$ the system
is in the gapless Ising ferromagnet universality class. In the critical
regime, the spin-spin correlation function decays as a power law with
well-known exponents~\cite{luther-peschel} and conjectured-exact
amplitudes~\cite{affleck-lukyanov}. In addition, it was shown that
the pairwise entanglement between nearest-neighbor spins is maximum
at the isotropic point~\cite{gu,syljuasen}. However, the pairwise
entanglement between further neighbors has a maximum near $\Delta=-1$~\cite{syljuasen}.
Moreover, the ground-state pairwise entanglement is not affected by
spontaneous symmetry breaking~\cite{syljuasen}.

\begin{figure}[h]
\begin{center}\includegraphics[%
  clip,
  width=0.85\columnwidth,
  keepaspectratio]{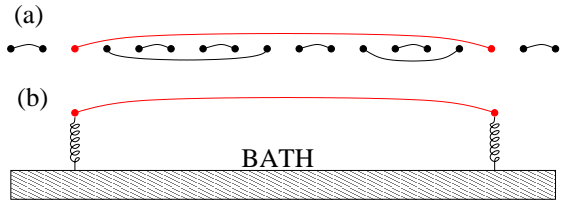}\end{center}

\caption{(Color online) (a) Pictorial view of the random singlet ground state.
Bonds connecting spins denote singlet pairs. (b) The ground state
from the point of view of the highlighted spin pair.\label{cap:ground-state}}
\end{figure}

In contrast to the clean model, the low-energy physical behavior of
all the chains parametrized by Hamiltonian (\ref{eq:hamiltonian})
in which $-1/2<\Delta_{i}\leq1$ is completely different, no matter
how weak the disorder strength is. In the renormalization-group (RG)
sense, it is said that the chains are governed by the so-called IRFP~\cite{fisher94-xxz}.
Within this framework, their ground states are a collection of spin
singlets whose spin pairs can be arbitrarily faraway from each other
{[}see Fig.~\ref{cap:ground-state}(a){]}. %

It is also known that the chance of finding a very distant spin singlet
pair is rare---viz., $\sim\left|i-j\right|^{-2}$. However, these
rare spin pairs contribute enormously to the \emph{average} spin-spin
correlation function, since they develop correlations of order unity.
Hence, the \emph{mean} correlation function decays algebraically as
$\left|i-j\right|^{-2}$~\cite{fisher94-xxz}. In contrast, a \emph{typical}
faraway spin pair is not in a singlet state, developing weak correlation
$\sim\exp\{-\sqrt{\left|i-j\right|}\}$, meaning a \emph{typical}
correlation function decaying as $\exp\{-\sqrt{\left|i-j\right|}\}$~\cite{fisher94-xxz}. 

How close those rare faraway singlets are to a perfect singlet (or
Bell) state is the question we address in this work. Moreover, we
wish to quantify their entanglement. As we show in Sec.~\ref{sec:Numerical-results},
there are rare and extreme faraway spin pairs displaying enormous
fidelity and entanglement. This is a remarkable property absent in
the clean and ordered systems: Two spins separated by a distance of
order of the chain length and interacting indirectly through the spins
between them are strongly entangled and, therefore, weakly entangled
with the rest of the chain. 

We now explain how those rare spin pairs are located in the chain.
Given a disordered chain they can be easily found using the decimation
rules predicted by the strong-disorder renormalization group method~\cite{MDH,fisher94-xxz}.
One has to search for the greatest coupling constant in the chain---say,
$J_{2}$---and then decimate spins $S_{2}$ and $S_{3}$ by coupling
them in a singlet state. The neighbors spins $S_{1}$ and $S_{4}$
now interact with each other through a new renormalized coupling constant
$\tilde{J}=J_{1}J_{3}/\left[\left(1+\Delta_{2}\right)J_{2}\right]$.
The local anisotropic parameter also renormalizes to $\tilde{\Delta}=\Delta_{1}\Delta_{3}\left(1+\Delta_{2}\right)/2$.
Iterating this procedure one can find the position of all spin pairs
in the chain. Since all the coupling constants (including the renormalized
ones) are random independent variables, the singlet pairs are distributed
in a random fashion, and thus the system is said to be in a random
singlet phase.

\subsection{Entanglement properties}

Having localized the spin pairs we now want to study how close they
are to a pure singlet---i.e., the Bell state $|\Psi^{-}\rangle=\left(\left|+-\right\rangle -\left|-+\right\rangle \right)/\sqrt{2},$
with $|+\rangle$ and $|-\rangle$ meaning spin up and down, respectively.
For this purpose we calculate the fidelity $F_{ij}$ between the actual
state describing the random singlets $\rho_{ij}$ and $|\Psi^{-}\rangle$,
which is given as \begin{equation}
F_{ij}=\left\langle \Psi^{-}\left|\rho_{ij}\right|\Psi^{-}\right\rangle ,\label{eq:F}\end{equation}
 where $\rho_{ij}=\textrm{tr}_{\overline{ij}}(\rho)$ is the usual
reduced density matrix of spins $S_{i}$ and $S_{j}$ (obtained tracing
out all spins but $S_{i}$ and $S_{j}$) and $\rho=\left|\mathcal{G}\right\rangle \left\langle \mathcal{G}\right|$
is the zero temperature density matrix in which $\left|\mathcal{G}\right\rangle $
is the ground state of the Hamiltonian (\ref{eq:hamiltonian}). Note
that $F_{ij}=1$ if $\rho_{ij}$ is a perfect singlet and that $F_{ij}=0$
whenever it is orthogonal to $|\Psi^{-}\rangle$. 

Using symmetry arguments we can show that $\rho_{ij}=4\sum_{a}C_{ij}^{aa}S_{i}^{a}S_{j}^{a}+\mathbf{1}/4$,
with $a=x,\, y,\, z$ and $C_{ij}^{aa}=\left\langle \mathcal{G}\left|S_{i}^{a}S_{j}^{a}\right|\mathcal{G}\right\rangle $
being the ground state spin-spin correlation. In addition, $C_{ij}^{xx}=C_{ij}^{yy}$
for the XXZ AF spin-1/2 chain. Expanding $\rho_{ij}$ in the Bell
basis, $|\Phi^{\pm}\rangle=(|++\rangle\pm|--\rangle)/\sqrt{2}$ and
$|\Psi^{\pm}\rangle=(|+-\rangle\pm|-+\rangle)/\sqrt{2}$, we get \begin{eqnarray}
\rho_{ij} & = & F_{ij}|\Psi^{-}\rangle\langle\Psi^{-}|+\left(4C_{ij}^{xx}+F_{ij}\right)|\Psi^{+}\rangle\langle\Psi^{+}|\nonumber \\
 &  & +\left(\frac{1}{4}+C_{ij}^{zz}\right)\left(|\Phi^{+}\rangle\langle\Phi^{+}|+|\Phi^{-}\rangle\langle\Phi^{-}|\right),\label{eq:decomposition}\end{eqnarray}
 where \begin{equation}
F_{ij}=\frac{1}{4}-2C_{ij}^{xx}-C_{ij}^{zz}.\label{eq:fidelity}\end{equation}

However, we are also interested in a complete quantification of the
entanglement between spins $S_{i}$ and $S_{j}$. Therefore, we compute
its negativity~\cite{werner} which can be defined as $\mathcal{N}_{ij}=||\rho_{ij}^{T_{i}}||-1$,
where $||A||=\textrm{tr}(\sqrt{A^{\dagger}A})$ is the trace norm
of operator $A$ and $T_{i}$ means partial transposition, in the
sense that the transposition operation is performed only in spin $S_{i}$~\cite{peres}.
The normalization for $\mathcal{N}_{ij}$ was chosen such that it
is equal to 1 for a maximally entangled state (e.g., Bell state).
A direct calculation using Eq.~(\ref{eq:decomposition}) gives \begin{equation}
\mathcal{N}_{ij}=\left\{ \begin{array}{ll}
0, & \text{if}\,\, F_{ij}\leq1/2,\\
2F_{ij}-1, & \text{if}\,\, F_{ij}>1/2.\end{array}\right.\label{eq:negativity}\end{equation}
 We remark that whenever $F_{ij}>1/2$ the two spins are entangled
and that the greater the fidelity the greater the entanglement. The
entanglement threshold $F_{ij}=1/2$ will be an important quantity
in the discussions below. Note that the negativity is a function of
the fidelity, implying that knowledge of the latter allows us to obtain
the negativity through Eq.~(\ref{eq:negativity}). 

Moreover, a complete characterization of the entanglement properties
is desirable. Interestingly, for the models described by the Hamiltonian
(\ref{eq:hamiltonian}), we find that the logarithmic negativity ${\mathcal{E}}_{ij}$~\cite{werner}
and the concurrence ${\mathcal{C}}_{ij}$~\cite{concurrence} can
be expressed in terms of the fidelity: \begin{equation}
{\mathcal{E}}_{ij}=\log_{2}\left|\left|\rho_{ij}^{T_{i}}\right|\right|=\left\{ \begin{array}{ll}
0, & \text{if}\,\, F_{ij}\leq1/2,\\
\log_{2}\left(2F_{ij}\right), & \text{if}\,\, F_{ij}>1/2,\end{array}\right.\label{eq:lognegat}\end{equation}
 and \begin{equation}
{\mathcal{C}}_{ij}=\max\left\{ 0,\lambda_{1}-\lambda_{2}-\lambda_{3}-\lambda_{4}\right\} ={\mathcal{N}}_{ij},\label{eq:concurrence}\end{equation}
 where the $\lambda$'s are the square roots of the eigenvalues of
the matrix $\rho_{ij}\left(\sigma_{i}^{y}\otimes\sigma_{j}^{y}\right)\rho_{ij}^{*}\left(\sigma_{i}^{y}\otimes\sigma_{j}^{y}\right)$
arranged in decreasing order~\cite{footnote2}, $\sigma^{y}$ is
the usual Pauli matrix, and $\rho^{*}$ is the complex conjugate of
$\rho$. The logarithmic negativity is an additive entanglement measure
for mixed states not recovering the von Neumann entropy for pure states
in general~\cite{werner}. However, the entanglement of formation~\cite{concurrence},
${\mathcal{E}}_{f}=-x\log_{2}x-\left(1-x\right)\log_{2}\left(1-x\right)$,
with $x=1/2+\sqrt{1-{\mathcal{C}}^{2}}/2$, is a monotonous function
of the concurrence and does reduce to the von Neumann entropy for
pure states. 

We thus conclude that, due to the symmetries of the Hamiltonian (\ref{eq:hamiltonian}),
knowledge of the fidelity is sufficient to completely quantify the
entanglement between any pair of qubits in the chain.

\section{Numerical results\label{sec:Numerical-results}}

We now report our numerical study on the fidelity between spin pairs.
Here we employed chains with coupling constants distributed according
to \begin{equation}
P\left(J\right)=\frac{\left(1-\alpha\right)}{\Omega_{0}^{1-\alpha}}J^{-\alpha},\label{eq:P(J)}\end{equation}
 where $0<J<\Omega_{0}$ and the disorder parameter $\alpha<1$. Hence,
greater $\alpha$ means greater disorder and $\Omega_{0}$ is the
energy scale~\cite{footnote3}. We have considered three different
disorder parameters $\alpha=0,$ $\alpha=0.3$, and $\alpha=0.6$
and chains of length $L_{0}=100$, $200$, $400$, and $800$ with
periodic boundary conditions. 

\begin{figure}[h]
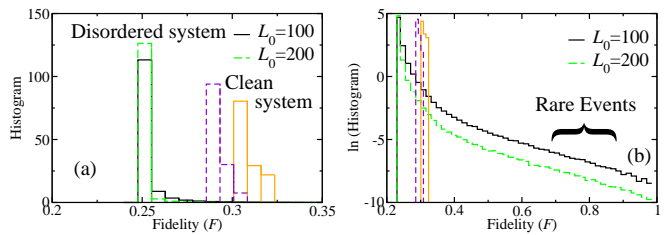

\begin{center}\includegraphics[%
  clip,
  width=0.9\columnwidth,
  height=3cm,
  keepaspectratio]{fig2a.eps}~ \includegraphics[%
  clip,
  height=3cm,
  keepaspectratio]{fig2b.eps}\end{center}

\caption{(Color online) Normalized histogram of fidelities for those spin
pairs separated by distances greater than a cutoff $l_{c}=L_{0}/6$
for the disordered {[}$\alpha=0.3$, leftmost solid (black) and dashed
(green) curves{]} and the clean and ordered chains {[}rightmost solid
(orange) and dashed (purple) curves{]}, and for length chains $L_{0}=100$
(solid lines) and $L_{0}=200$ (dashed lines). (a) The typical value
of $F$ is clearly greater in the clean chain than in the disordered
one. (b) However, as revealed by the logarithmic scale, there are
some rare spin pairs (rare events) displaying high fidelities in the
disordered chain, while in the clean one the fidelities are sharply
constrained to values less than 1/2. A total of $6\,000$ different
disordered chains were used to build each one of these histograms.\label{cap:Histogram}}
\end{figure}

In order to emphasize the difference between the clean and disordered
systems, we plot in Fig.~\ref{cap:Histogram} the histogram of fidelities
for all spin pairs separated by odd lengths greater than a cutoff
$l_{c}=L_{0}/6$~\cite{footnote4}.%
As shown in panel (a), the fidelities of those spins in the disordered
chain are \emph{typically} less than those in the ordered one. This
is because in the disordered system the typical correlation function
decays $\sim\exp\{-\sqrt{\left|i-j\right|}\}$; hence, the \emph{typical}
value of $F_{ij}$ approaches $1/4$ rapidly when $L_{0}$ is increased,
while in the clean system the transversal and longitudinal correlation
function decays as $\left|i-j\right|^{-\eta}$ with $\eta=1/2$ and
2, respectively~\cite{footnote5}. Moreover, as $C_{ij}^{aa}$ depends
only on $\left|i-j\right|$ in the ordered system, the fidelities
are thus sharply bounded as dictated by Eq.~(\ref{eq:fidelity})
and clearly shown in panel (b). As also shown in panel (b), on the
other hand, the fidelities in the disordered chain are not sharply
bounded from above. It is due to the fact that $C_{ij}^{aa}$ is distributed
for a fixed $\left|i-j\right|$. It is worth mentioning that there
exist some rare spin pairs displaying strikingly high fidelities.
Although they are very rare, their fidelities are above the threshold
$1/2$ below which there is no entanglement {[}see Eqs.~(\ref{eq:negativity})-(\ref{eq:concurrence}){]}.
In contrast, there is no distant spin pair displaying such high fidelity
in the clean chain. 

We now focus our attention on those rare and distant pairs. In order
to characterize their entanglement, we choose one spin pair per chain
to represent them. There are two natural choices: (i) the longest
and (ii) the last decimated spin pairs. The first choice is interesting
because we are focusing on pairs whose spins are extremely distant.
However, we choose the second one by the following reasons: (i) it
is easier and faster to implement numerically, (ii) there is a great
chance the last decimated pair be the longest one, and (iii) as we
shall see below, we focus our attention only on pairs separated by
distances of order of the chain length. We thus proceed as follows. 

\begin{figure}[h]
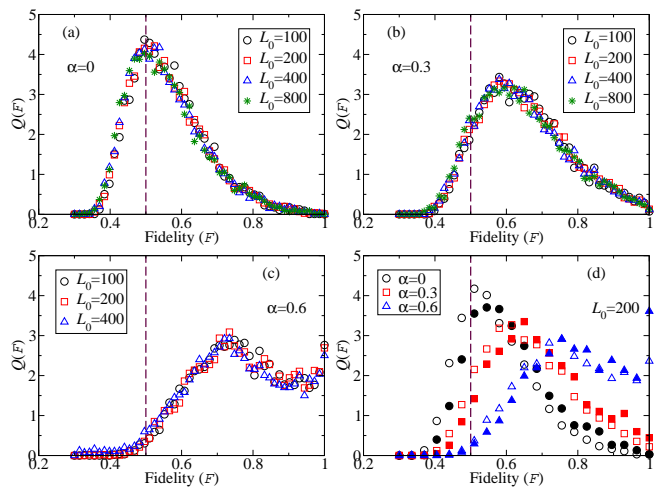

\begin{center}\includegraphics[%
  clip,
  width=0.49\columnwidth,
  keepaspectratio]{fig3a.eps}~\includegraphics[%
  clip,
  width=0.49\columnwidth,
  keepaspectratio]{fig3b.eps}\\
\includegraphics[%
  clip,
  width=0.49\columnwidth,
  keepaspectratio]{fig3c.eps}~\includegraphics[%
  clip,
  width=0.49\columnwidth,
  keepaspectratio]{fig3d.eps}\end{center}

\caption{(Color online) Fidelity distribution $Q\left(F\right)$ for the chains
with disorder parameter (a) $\alpha=0$, (b) $\alpha=0.3$, and (c)
$\alpha=0.6$, concerning the last decimated spin pairs separated
by distances greater than $l_{c}=L_{0}/6$. The same in panel (d)
for a fixed length $L_{0}=200$ and different disorder parameters.
Open (solid) symbols refer to spin pairs separated by distances greater
(shorter) than $l_{c}$. The vertical dashed line highlights the threshold
$F=1/2$ below which there is no entanglement. \label{cap:Distribution-of-fidelity}}
\end{figure}

We entirely decimate a chain following the RG rules~\cite{MDH} in
order to find the position of the last spin pair. Then, we compute
their spin-spin correlations exactly~\cite{lieb-schultz-mattis}.
With this information we calculate the fidelity between this last
pair and a singlet as given in Eq.~(\ref{eq:fidelity}). In Fig.~\ref{cap:Distribution-of-fidelity}
we plot the fidelity distribution $Q\left(F\right)$ for chains with
$\alpha=0$, $0.3$, and $0.6$ {[}see panels (a), (b), and (c), respectively{]}
for different chain lengths. For each length and disorder parameter,
we computed a total of $10\,000$ fidelities in order to build the
distributions $Q\left(F\right)$. In panels (a)--(c) we retained only
those pairs separated by a distance between $L_{0}/6$ and $L_{0}/2$
(around 70\% of the last spin pairs follow this requirement, independent
of disorder parameter and chain length). In all the cases studied,
there is no difference between the histograms for different chain
lengths. \textit{\emph{We thus conclude that the probability of finding}}
\textit{faraway and highly entangled spin pairs} \textit{\emph{does
not depend on the chain length and increases with the disorder strength.}}%

Moreover, looking at panels (a)--(c) and especially panel (d) of Fig.~\ref{cap:Distribution-of-fidelity}
we clearly see that an increasing disorder increases the probability
of obtaining a spin pair with greater fidelity and greater entanglement.
We also highlight that for greater disorder {[}panel (c){]} we have
a reasonable probability of finding pairs of spins with $F_{ij}\approx1$.
Those pairs can be considered, for all practical purposes, perfect
singlets. This is in contrast with ordered and clean chains in which
pairwise entanglement is found only between nearest neighbors~\cite{syljuasen,osborne,osterloh}. 

We want now to draw some attention to the following caveat. The mean
value of the fidelities concerning all spin pairs separated by distances
of order of the chain length approaches 1/4 as the chain length increases.
However, there are always some rare spin pairs displaying enormous
entanglement {[}as shown in Fig.~\ref{cap:Histogram}(b){]}. Moreover,
the probability of finding them is $\sim\left|i-j\right|^{-2}$; i.e.,
it depends only on their separation~\cite{fisher94-xxz}. Precisely,
for $\left|i-j\right|\gg1$, the probabity of finding them equals
$2/(3\left|i-j\right|^{2})$~\cite{hoyos-correlation}. Furthermore,
the fidelity distribution concerning only those rare spin pairs $Q\left(F\right)$
is completely different. It does \emph{not} depend on the separation
between them, and its mean value increases with the disorder strength
{[}as shown in Figs.~\ref{cap:Distribution-of-fidelity}(a)--\ref{cap:Distribution-of-fidelity}(c){]}.

Our results are in agreement with the localized nature of the ground
state. Spin pairs separated by distances greater than the corresponding
order-disorder crossover length shall share correlations mediated
by virtual polarizations of the spin singlets between them~\cite{footnote6}.
Note that the fidelity distributions of those last pairs separated
by distances shorter than $L_{0}/6$ is slightly shifted to the right
{[}see Fig.~\ref{cap:Distribution-of-fidelity}(d){]}; i.e., the
mean value of the fidelity is greater. This result does not mean there
is a length dependence on $Q\left(F\right)$ since there are many
other short-length spin pairs earlier decimated without such high
fidelity. Moreover, there are pairs separated by distances less than
the ordered-disordered crossover length, meaning we cannot simply
compare such distributions.

Besides the existence of a reasonable number of almost pure singlets
for strong disorder, we also have a remarkable ensemble of states
suited for the implementation of distillation protocols~\cite{bennett-dist}.
A distillation protocol can be seen as a local procedure by which
we can obtain $M$ singlets from an ensemble of $N>M$ mixed states.
One can show~\cite{bennett-dist} that only mixed states $\rho_{ij}$
with a fidelity, as given by Eq.~(\ref{eq:F}), greater than 1/2
can be employed for distillation purposes. And this is exactly the
scenario we have in random AF spin-1/2 chains. For example, for a
disorder parameter $\alpha=0.3$ we have almost $90\%$ of the last
decimated singlets with $F_{ij}>1/2$ and we can find many pairs of
spins more than $100$ sites apart with fidelities above this threshold.
For $\alpha=0.6$ we achieve rates greater than $97\%$. Remarkably,
those spins are extremely distant from each other---namely, fractions
of the chain size. 

Typically, the effective coupling $J_{eff}$ between distant spins
is exponentially weak---viz., $\Omega_{0}\exp\{-\gamma\sqrt{\left|i-j\right|}\}$,
where $\gamma$ is a constant proportional to the disorder strength
but of order of unit~\cite{fisher94-xxz}. Analyzing the distribution
of $J_{eff}$ we find $\gamma\approx1.2$, $2.0$, and $3.5$ for
$\alpha=0$, $0.3$, and $0.6$, respectively. Thus, the typical value
of $J_{eff}$ with respect to a chain with disorder parameter $\alpha=0.6$
for a pair separated by a distance, say, $\left|i-j\right|=40$ is
of order $10^{-10}\Omega_{0}$. With this result, one may be concerned
about effects of local perturbations on the singlets and also effects
of finite temperature. However, although the typical value of $J_{eff}$
decreases $\sim\exp\{-\sqrt{\left|i-j\right|}\}$, the width of the
distribution of $\ln(J_{eff})$ increases $\sim\sqrt{\left|i-j\right|}$,
meaning there is a finite probability to find a high-fidelity spin
singlet with relatively strong coupling. Among our data regarding
chains with lengths up to 100 sites and $\alpha=0.6$, we found 5
chains where the distances between the spins were greater than 40
sites, the fidelities were greater than 0.75, and their effective
couplings were greater than $10^{-6}\Omega_{0}$. From the energetic
point of view, less disordered chains are favorable. For the chain
with $\alpha=0$ and same length, we found 11 of those spin pairs
but their effective coupling were greater than $10^{-4}\Omega_{0}$.

\section{Further discussions and conclusion\label{sec:Conclusions}}

We now put our results in a more general context. Consider first the
random AF dimerized spin chains whose Hamiltonian is given by Eq.~(\ref{eq:hamiltonian})
with different coupling constant distributions between odd and even
bonds. In the Griffiths phase near criticality (i.e., the even and
odd coupling constant distributions are not identical but have great
overlap), the strong-disorder RG scheme also applies and spin pairs
couple in singlet states~\cite{hyman-dimer}. Thus, random spin pairs
resembling perfect singlets are also expected. However, the probability
of finding distant spins in a singlet state is lower when compared
to the model studied here. Also, it is known that the random AF two-leg
and zigzag spin-1/2 ladders renormalize to random critical or dimerized
spin chains in certain conditions~\cite{hoyos-ladders}. This suggest
that the results here presented also apply to those systems. 

Let us now raise an interesting question: if the singletlike spin
pairs are weakly perturbed, how do they behave after a projective
measurement? It is reasonable to expect they will oscillate coherently
with a period inversely proportional to their effective coupling constant.
This means that spins $\left|i-j\right|$ sites away will slowly oscillate
with period $\sim\exp\{\sqrt{\left|i-j\right|}\}$ while the rest
of the chain stays unperturbed~\cite{footnote7}. Note that in this
strong-disorder scenario there is no room for the characteristic {}``decoherence
wave'' of the clean system~\cite{hamieh-katsnelson}. The mechanism
behind such {}``protection'' against decoherence is certainly related
with the localized aspect of the system, where the effective coupling
between the spin pair and the rest of the chain is weak {[}see Fig.~\ref{cap:ground-state}(b){]}. 

In conclusion, we have shown that the ground state of random AF spin-1/2
chains displays remarkable entanglement properties. First, it is possible
to find long-distance spin pairs that are almost perfect singlets.
Actually, the more disorder the system possesses, the greater is the
probability of finding such pairs, meaning that they become weakly
perturbed singletlike states. These results are in agreement with
what is expected from the localized nature of the ground state as
given by the strong-disorder RG scenario of the random singlet phase.
However, the fidelities and negativities here calculated provided
the first suitable and exact quantification of how much such pairs
resembles a singlet state. Moreover, we have shown that the distribution
of the fidelity and entanglement between distant singlet-like pairs
does not depend on the chain length, depending only on the disorder.
Consequently, the probability to find a high-fidelity spin pair is
equal to $2Q\left(F\right)/(3\left|i-j\right|^{2})$, provided the
distance $\left|i-j\right|$ between them be much longer than the
corresponding order-disorder crossover length. Notice the only dependence
on disorder is given in $Q\left(F\right)$ as shown in Figs.~\ref{cap:Distribution-of-fidelity}(a)--\ref{cap:Distribution-of-fidelity}(c).
In addition to showing that some spin pairs are very good quantum
channels (fidelity with a singlet $\approx1$) we have shown that
all the weakly perturbed spin pairs can be used to create perfect
singlets. For strong disorder we have shown that more than $90\%$
of these pairs possess fidelities greater than the lower bound 1/2
above which quantum distillation protocols, procedures in which a
few perfect singlets can be obtained from a large ensemble of mixed
states, become feasible. 

We also point out that the use of local defects in related models
in order to achieve strongly entangled states was reported in Ref.~\cite{santos}.
However, the entanglement there were obtained between near spins (qubits)
in contrast to those here reported. 

Finally, it has been recently achieved an experimental realization
of the AF spin-5/2 Heisenberg model~\cite{hirjibehedin}. The astonishing
experimental precision achieved by the scanning tunneling microscope
leads to an unprecedent experimental possibility of confronting the
theoretical predictions of the effects of disorder on low-dimensional
magnets. In particular, this experiment suggests that in the near
future the results here presented can be tested and applied to the
creation of highly entangled quantum channels. 

\begin{acknowledgments}
J.A.H. would like to thank T. Vojta for useful discussions and acknowledges
support from NSF under Grant No. DMR-0339147 and from Research Corporation.
G.R. is supported by Funda\c{c}\~ao de Amparo \`a Pesquisa do Estado
de S\~ao Paulo (FAPESP). 
\end{acknowledgments}


\begin{thebibliography}{17}
\bibitem{footnote1}Given a spin chain we can partition it into two complementary blocks
$A$ and $B$. The bipartite entanglement entropy is the von Neumann
entropy of either subsystem $A$ or $B$.
\bibitem{footnote2}The assumptions yielding Eqs.~(\ref{eq:negativity})--(\ref{eq:concurrence})
are (i) $C_{ij}^{aa}$ are between zero and 1/4 if $\left|i-j\right|$
is even and between $-1/4$ and zero, otherwise; and (ii) $\rho_{ij}$
has non-negative eigenvalues.
\bibitem{footnote3}An appropriate disorder measurer is the ratio between the width of
the distribution and its mean---i.e., $\sqrt{{\textrm{Var}}J}/\langle J\rangle$,
where Var$J$ is the variance of $J$. In the limit $\alpha\rightarrow1$
the disorder strength diverges.
\bibitem{footnote4}The negativity between spin pairs separated by even distances is zero;
we thus neglect them since their entanglement properties are not interesting.
\bibitem{footnote5}Note that one can easily compute the histogram of the fidelities,
since the behavior of the correlation functions are known. For the
clean system, it behaves $\sim\left(F-1/4\right)^{-\left(1+1/\eta\right)}$,
where $\eta=1/2$ since $F$ is dominated by the transversal correlation
function. For the disordered system, the histogram diverges $\sim\ln\left(F-1/4\right)/\left(F-1/4\right)$
as $F\rightarrow1/4$.
\end{thebibliography}
\end{document}